\documentclass[a4paper]{article}
\title{
\vspace{-3mm}
\rightline{\small IFUP-TH 2003/25}
\vspace{8mm}
\bf Heavy-quark condensate at zero- and nonzero temperatures for
various forms of the short-distance potential} 
\author{Dmitri Antonov \thanks{
E-mail address: {\tt antonov@df.unipi.it}}~ \thanks{Permanent address:
ITEP, B. Cheremushkinskaya 25, RU-117 218 Moscow, Russia.}\\
{\it INFN-Sezione di Pisa, Universit\'a degli studi di Pisa,
Dipartimento di Fisica ``E. Fermi'',}\\
{\it Via Buonarroti, 2 - Ed. B-C -
I-56127 Pisa, Italy}}

\date{}
\begin{document}

\maketitle
\vspace{1mm}
\centerline{\bf {Abstract}}
\vspace{3mm}
\noindent
With the use of the world-line formalism, 
the heavy-quark condensate in the SU(N)-QCD is evaluated for the cases when the 
next-to-$1/r$ term in the quark-antiquark potential at short distances is either 
quadratic, or linear. In the former case, the standard QCD-sum-rules result 
is reproduced, while the latter result is a novel one.
Explicitly, it is UV-finite only in less than four
dimensions. This fact excludes a possibility to have, in four dimensions, very short strings 
(whose length has the scale of the lattice 
spacing), and consequently the short-range linear potential (if it exists) cannot violate
the OPE. In any number of dimensions, the obtained novel expression for the quark condensate 
depends on the 
string tension at short distances, rather than on the gluon condensate, and grows linearly
with the number of colors in the same way as the standard QCD-sum-rules expression.
The use of the world-line formalism enables one to generalize further both results to the 
case of finite temperatures. A generalization of the QCD-sum-rules expression to the case of 
an arbitrary number of space-time dimensions
is also obtained and is shown to be UV-finite, provided this number is smaller than six.

\vspace{10mm}

\section{Introduction.}
One possible way to evaluate the quark condensate is to deduce it from the (one-loop) 
quark self-energy (i.e., the averaged one-loop quark
effective action), since these quantities are related to each other by
the following formula:

\begin{equation}
\label{1}
\left< \bar \psi \psi \right>=-\frac{1}{V}
\frac{\partial}{\partial m} \left< \Gamma[A^a_\mu]
\right>.
\end{equation}
Here,  $V$ is the four-volume
occupied by the system, $m$ is the current quark mass, the average
$\left<\ldots\right>$ is defined with respect to the gluodynamics action
in the Euclidean space-time, $\frac14 \int d^4 x\left(F^a_{\mu\nu}\right)^2$,
where $a=1,\ldots, N^2-1$ and $F^a_{\mu\nu} =\partial _\mu
A^a_\nu-\partial_\nu A_\mu^a+gf^{abc} A_\mu^bA_\nu^c$ stands for
the YM field-strength tensor. For the spin-$\frac12$ quarks, the 
one-loop quark self-energy $\left<
\Gamma[A^a_\mu]\right>$ reads~\cite{1} (see~\cite{2} for a
recent review):

$$
\left<\Gamma\left[A_\mu^a\right]\right>=
-2\int\limits_{\Lambda^{-2}}^{\infty}\frac{dT}{T}{\rm e}^{-m^2T}
\int\limits_{P}^{} {\cal D}x_\mu
\int\limits_{A}^{} {\cal D}\psi_\mu
\exp\left[-\int\limits_{0}^{T}d\tau\left(\frac14\dot x_\mu^2+
\frac12\psi_\mu\dot\psi_\mu\right)\right]\times$$

\begin{equation}
\label{2}
\times\left\{\left<{\rm tr}{\,}{\cal P}\exp\left[
ig\int\limits_{0}^{T}d\tau\left(A_\mu\dot x_\mu-\psi_\mu\psi_\nu
F_{\mu\nu}\right)\right]\right>-N\right\}.
\end{equation}
Here, $\Lambda$ stands for the UV momentum cutoff, the subscripts $P$ and $A$ imply the periodic and antiperiodic 
boundary conditions of the respective path-integrals, 
$\psi_\mu$'s are antiperiodic Grassmann functions (superpartners of $x_\mu$'s), and 
$A_\mu\equiv A_\mu^a T^a$ with $T^a$'s standing for the 
generators of the SU(N)-group in the fundamental 
representation, $\left[T^a,T^b\right]=if^{abc}T^c$,
${\rm tr}~ T^aT^b=\frac12\delta^{ab}$.~\footnote{Note that 
the sign ``$+$'' at ``$ig$'' in eq.~(\ref{2}) differs from that of ref.~\cite{2} and stems 
from the definition of the covariant derivative $D_\mu=\partial_\mu-igA_\mu^aT^a$, that itself is unambiguously related to the 
above-presented definition of $F_{\mu\nu}^a$.}
We have also adopted the normalization 
conditions $\left<\Gamma[0]\right>=0$ and 
$\left<W(0)\right>=N$, where 

\begin{equation}
\label{3}
\left<W({\cal C})\right>\equiv
\left<{\rm tr}{\,}{\cal P}\exp\left(ig\int\limits_{0}^{T}
d\tau A_\mu\dot x_\mu\right)\right>
\end{equation}
is the Wilson loop with the contour ${\cal C}$ parametrized 
by the vector-function $x_\mu(\tau)$.

To evaluate the path integral~(\ref{2}), a certain Ansatz for $\left<W({\cal C})\right>$ should be implemented~\footnote{In what follows, we 
mean by $\left<W({\cal C})\right>$ that part of the full Wilson loop, which generates terms starting from the linear one in the quark-antiquark potential.
The Coulomb term yields the multiplicative renormalization of the loop (see e.g. ref.~\cite{4}), 
and the so-appearing renormalization factor is supposed to be 
included in the definition of the path-integral measure in eq.~(\ref{2}). The same concerns other possible (nonperturbative) 
$1/r$-terms in the potential, e.g. the L\"uscher term.}. 
In ref.~\cite{testing}, the Vacuum Correlator Method (VCM)~\cite{55, 6} (for a recent review see~\cite{5})
has been used for a parametrization of the 
Wilson loop. This Ansatz enabled one to find a relation between the short-distance asymptotics of the two functions, which parametrized
the two-point gauge-invariant correlation function of the YM field strengths within the VCM. The aim of the present letter is to explore the 
heavy-quark condensate by the same world-line method, but for various forms of the next-to-$1/r$ term in the short-distance quark-antiquark 
potential. While VCM predicts this form to be quadratic~\cite{6} (that, as we shall see, yields the QCD-sum-rules expression~\cite{7} 
for the heavy-quark condensate), numerous discussions have recently been made in favor of the 
linear form~\cite{10, yuas, hrs}. A naturally arising question is then: what is the expression for the heavy-quark condensate in the case 
when the leading next-to-$1/r$ term in the short-distance potential is linear? An answer to this question for an arbitrary number of 
space-time dimensions will be obtained in the next section. Besides the direct physical application in 2d and 3d (the latter of which
will also be important for the finite-temperature analysis), an interest to the 
generalization to an arbitrary number of dimensions is that the condensate turns out to be logarithmically UV divergent in 4d.
This result means that the short-string thickness (that plays the role of the inverse UV cutoff) cannot be as small as the lattice 
spacing, but should rather be comparable with the vacuum correlation length. In another words, had strings with a length comparable with  
the lattice spacing existed, this would lead to an UV divergency of the heavy-quark condensate, that cannot be the case. 
In the language of an effective gluon mass~\cite{yuas, hrs} this fact means that a tachyonic gluon mass vanishes 
in the UV regime, and OPE is not violated by short strings (since the latter cannot be ``very short'').

The letter is organized as follows.  
In the next section, we shall start with the evaluation of the heavy-quark condensate in four and further in arbitrary
number of dimensions, for the quadratic form of the next-to-$1/r$ term in the short-distance potential~\footnote{Clearly, in 4d, the 
QCD-sum-rules result will be reproduced.}. In section 3, the analysis in an arbitrary number of dimensions
will be extended to the case of the linear form of the short-distance potential.
In section 4, the finite-temperature generalizations of both results will be derived in 4d.
Finally, in Summary section, all the obtained results will be summarized.

\section{The heavy-quark condensate in the case 
of quadratic form of the next-to-$1/r$ term in the short-distance potential.}
Let us start with the analysis of eq.~(\ref{2}) in 4d. 
As it has been discussed in ref.~\cite{testing}, all 
the gauge-field dependence of eq.~(\ref{2}) can be reduced 
to that of the Wilson loop~(\ref{3}).  
That is because (as it has been shown e.g. in ref.~\cite{3}) 
the spin part of the world-line action 
can be rewritten by means of the operator of the area derivative 
of the Wilson loop as follows:

\begin{equation}
\label{4}
\left<{\rm tr}{\,}{\cal P}\exp\left[
ig\int\limits_{0}^{T}d\tau\left(A_\mu\dot x_\mu-\psi_\mu\psi_\nu
F_{\mu\nu}\right)\right]\right>=
\exp\left(-2\int\limits_{0}^{T}d\tau\psi_\mu\psi_\nu
\frac{\delta}{\delta\sigma_{\mu\nu}(x(\tau))}\right)
\left<W({\cal C})\right>.
\end{equation}
Next, the volume factor $V$ becomes extracted from eq.~(\ref{2}) by 
splitting the coordinate $x_\mu(\tau)$ into the center-of-mass 
and the relative coordinate~\cite{2} as
$x_\mu(\tau)=\bar x_\mu+z_\mu(\tau)$, where the center-of-mass 
is defined as $\bar x_\mu=\frac1T\int\limits_{0}^{T}
d\tau x_\mu(\tau)$. The empty integration over $\bar x$ then 
obviously yields the factor $V$.

Further, the Wilson loop~(\ref{3}) 
can be rewritten by virtue of the non-Abelian Stokes' theorem 
and the cumulant expansion as follows (see e.g.~\cite{5}):

\begin{equation}
\label{5}
\left<W({\cal C})\right>\simeq
{\,}{\rm tr}{\,}
\exp\left\{-\frac{1}{2!}\frac{g^2}{4}\int\limits_{\Sigma[{\cal C}]}^{}
d\sigma_{\mu\nu}(z)\int\limits_{\Sigma[{\cal C}]}^{}d\sigma_{\lambda\rho}(z')
\left<\left< 
F_{\mu\nu}(z)\Phi(z,z')F_{\lambda\rho}(z')
\Phi(z',z)\right>\right>\right\},
\end{equation}
where $\Phi(z,z')\equiv
{\cal P}{\,}\exp\left(ig\int\limits_{z'}^{z}
A_\mu(u)du_\mu\right)$ is the phase factor along the straight line, and 
$\left<\left<{\cal O}{\cal O}'\right>\right>\equiv
\left<{\cal O}{\cal O}'\right>-\left<{\cal O}\right>
\left<{\cal O}'\right>$ is the two-point irreducible average.
To get rid of the artificial $\Sigma$-dependence of the r.h.s. of eq.~(\ref{5}),
that is due to the neglection of irreducible averages higher than the quadratic one 
(the so-called bilocal or Gaussian approximation~\footnote{The symbol ``$\simeq$'' in eq.~(\ref{5})
is implied in the sense of this very approximation.}), $\Sigma$ is usually chosen to be the surface of the 
minimal area for a given contour ${\cal C}$ (see ref.~\cite{5} for discussions).
It is also worth commenting 
that the factor $1/2!$ in eq.~(\ref{5}) 
is simply due to the cumulant expansion, whereas the factor
$1/4$ is due to the (non-Abelian) Stokes' theorem with the 
usual agreement on the summation over {\it all} the indices (not only 
over those, among which the first one is smaller than the second, used 
in ref.~\cite{6}).

Next, the typical heavy-quark trajectories are small in a certain sense. Indeed, the 
free part of the bosonic sector of the 
world-line action reads

$${\cal S}_{\rm free}=
\frac14\int\limits_{0}^{T}d\tau\dot z_\mu^2(\tau)+m^2T=
\frac12\int\limits_{0}^{T}
dt\dot z_\mu^2(t)+\frac{m^2T}{2}.$$
Here, in the last equality we have performed a 
rescaling $\tau=\frac{t}{2}$, $T^{\rm old}=\frac{T^{\rm new}}{2}$.
Among all the reparametrization transformations of the contour, $t\to\sigma(t)$ with   
$\frac{d\sigma}{dt}\ge 0$,
let us choose the proper-time parametrization $t=\frac{s}{m}$.
Here, $s$ is the 
proper length of the contour, so that $T=L/m$, where  
$L=\int ds\equiv\int_{\sigma_{\rm in}}^{\sigma_{\rm fin}}d\sigma\sqrt{\dot
z_\mu^2(\sigma)}$ is the length of the contour. Within this 
parametrization we have  
$\int_{0}^{T}
dt\dot z_\mu^2(t)=m\int ds\left(\frac{dz_\mu(s)}{ds}\right)^2=
mL$, since $\left(\frac{dz_\mu(s)}{ds}\right)^2=1$ by the definition 
of the proper time. Therefore ${\cal S}_{\rm free}=mL$, implying  
that the typical heavy-quark trajectories are such that $L<{\cal O}(1/m)$.

For so small contours, we may use the following parametrization 
for the bilocal irreducible average (cumulant):

\begin{equation}
\label{cum4}
\left<\left< 
F_{\mu\nu}(z)\Phi(z,z')F_{\lambda\rho}(z')\Phi(z',z)
\right>\right>=
\frac{\hat 1_{N\times N}}{12N}{\rm tr}{\,}\left<F_{\mu\nu}^2\right>
\left(\delta_{\mu\lambda}\delta_{\nu\rho}-\delta_{\mu\rho}
\delta_{\nu\lambda}\right),
\end{equation}
and also approximate $\int_{\Sigma_{\rm min}}^{} d\sigma_{\mu\nu}(z)d\sigma_{\mu\nu}(z')=2S_{\rm min}^2$
by $\Sigma_{\mu\nu}^2$, where $\Sigma_{\mu\nu}=\oint z_\mu dz_\nu$ is the tensor area of the contour ${\cal C}$
(see e.g. ref.~\cite{4})~\footnote{Note that the equality of $S_{\rm min}$ to $\sqrt{\frac12\Sigma_{\mu\nu}^2}$
is exact only when ${\cal C}$ is a flat contour located in the $(\mu\nu)$-plane. However, being used for the evaluation of the 
path integral~(\ref{2}) in the case of small enough contours
(see the discussion above), this approximation will be proved to work perfectly, since in 4d it reproduces the 
QCD-sum-rules result.}.
The Wilson loop then takes the form~\cite{6}:

\begin{equation}
\label{quadr}
\left<W({\cal C})\right>\simeq N\exp\left(-C\Sigma_{\mu\nu}^2\right),
\end{equation} 
where $C\equiv\frac{g^2}{96N}\left<\left(F_{\mu\nu}^a(0)\right)^2\right>$.
The quadratic surface dependence of $-\ln\left<W({\cal C})\right>$ means that the leading next-to-$1/r$ term
in the short-distance quark-antiquark potential, one gets within the VCM, is also quadratic.

The intermediate expression for the heavy-quark self-energy~(\ref{2}) then reads:

$$\left<\Gamma\left[A_\mu^a\right]\right>=
-2NV\int\limits_{\Lambda^{-2}}^{\infty}\frac{dT}{T}{\rm e}^{-m^2T}
\int\limits_{P}^{} {\cal D}z_\mu
\int\limits_{A}^{} {\cal D}\psi_\mu
\exp\left[-\int\limits_{0}^{T}d\tau\left(\frac14\dot z_\mu^2+
\frac12\psi_\mu\dot\psi_\mu\right)\right]\times$$

\begin{equation}
\label{6}
\times
\left\{\exp\left(-2\int\limits_{0}^{T}d\tau\psi_\mu\psi_\nu
\frac{\delta}{\delta\sigma_{\mu\nu}(z(\tau))}\right)\exp\left(-C\Sigma_{\mu\nu}^2\right)-1\right\}.
\end{equation}
In order to proceed with the evaluation of this path integral, it is useful to linearize the 
$\Sigma_{\mu\nu}^2$-dependence by introducing the integration over an auxiliary antisymmetric 
tensor field. Since $\Sigma_{\mu\nu}$ depends on the contour as a whole, this field will crearly be 
space-time independent. Namely, we obtain:

$$
\exp\left(-C\Sigma_{\mu\nu}^2\right)=\frac{1}{(8\pi C)^3}
\left(\prod\limits_{\mu<\nu}^{}\int\limits_{-\infty}^{+\infty}dB_{\mu\nu}
{\rm e}^{-\frac{B_{\mu\nu}^2}{8C}}\right)
\exp\left(-\frac{i}{2}
B_{\mu\nu}\Sigma_{\mu\nu}\right).$$
Further, due to the (Abelian) Stokes' theorem
$B_{\mu\nu}\Sigma_{\mu\nu}=
\int_{\Sigma}^{}d\sigma_{\mu\nu}
B_{\mu\nu}$, and we have:

$$\exp\left(-2\int\limits_{0}^{T}d\tau\psi_\mu\psi_\nu
\frac{\delta}{\delta\sigma_{\mu\nu}(z(\tau))}\right)
{\rm e}^{-\frac{i}{2}B_{\mu\nu}\Sigma_{\mu\nu}}=
\exp\left(i\int\limits_{0}^{T}d\tau B_{\mu\nu}
\psi_\mu\psi_\nu-\frac{i}{2}B_{\mu\nu}\Sigma_{\mu\nu}\right).$$
We have therefore reduced the problem of evaluation of the heavy-quark self-energy in the YM field to the same 
problem for the electron in a constant field $B_{\mu\nu}$, which one should eventually average over.
Indeed, eq.~(\ref{6}) takes the form:

$$\left<\Gamma\left[A_\mu^a\right]\right>=
-\frac{2NV}{(8\pi C)^3}\int\limits_{\Lambda^{-2}}^{\infty}\frac{dT}{T}{\rm e}^{-m^2T}
\left(\prod\limits_{\mu<\nu}^{}\int\limits_{-\infty}^{+\infty}dB_{\mu\nu}
{\rm e}^{-\frac{B_{\mu\nu}^2}{8C}}\right)\times$$

\begin{equation}
\label{7}
\times\left\{
\int\limits_{P}^{} {\cal D}z_\mu
\int\limits_{A}^{} {\cal D}\psi_\mu
\exp\left[-\int\limits_{0}^{T}d\tau\left(\frac14\dot z_\mu^2+
\frac12\psi_\mu\dot\psi_\mu+\frac{i}{2}B_{\mu\nu}z_\mu\dot z_\nu
-iB_{\mu\nu}\psi_\mu\psi_\nu\right)\right]-\frac{1}{(4\pi T)^2}
\right\}.
\end{equation}
The expression in the curly brackets here is the famous Euler-Heisenberg-Schwinger Lagrangian
which reads (see e.g. refs.~\cite{2, 65}):

\begin{equation}
\label{8}
\left\{\ldots\right\}=\frac{1}{(4\pi T)^2}\left[T^2ab\cot(aT)\coth(bT)-1\right].
\end{equation}
Here, the standard notations have been used~\cite{2}:

$$a^2=\frac12\left[{\bf E}^2-{\bf H}^2+
\sqrt{\left({\bf E}^2-{\bf H}^2\right)^2+4({\bf E}
\cdot{\bf H})^2}{\,}\right],~~
b^2=\frac12\left[-\left({\bf E}^2-{\bf H}^2\right)+
\sqrt{\left({\bf E}^2-{\bf H}^2\right)^2+
4({\bf E}\cdot{\bf H})^2}{\,}\right]$$
with ${\bf E}=i\left(B_{41},B_{42},B_{43}\right)$ and ${\bf H}=\left(B_{23},-B_{13},B_{12}\right)$ 
[$B_{ij}=\varepsilon_{ijk}H_k$, $B_{4i}=-iE_i$]
being the electric and magnetic fields corresponding to the field-strength 
tensor $B_{\mu\nu}$.

Next, due to the factor ${\rm e}^{-m^2T}$ in eq.~(\ref{7}),
only small values of $T$ 
are sufficient in the heavy-quark limit, and eq.~(\ref{8})
can be expanded in powers of $T$. The leading term 
of this expansion, which in the expansion of the fermionic
determinant corresponds to the diagram 
with only two
external lines of the $B_{\mu\nu}$-field, stems from the following formula:

\begin{equation}
\label{approx}
T^2ab\cot(aT)\coth(bT)-1
=\frac{T^2}{3}\left(b^2-a^2\right)+
{\cal O}\left(T^4({\bf E}\cdot{\bf H})^2\right)=\frac{T^2}{3}\sum\limits_{\alpha<\beta}^{}B_{\alpha\beta}^2+{\cal O}
\left(T^4\left(B_{\mu\nu}^2\right)^2\right).
\end{equation}
Note that since $T\sim m^{-2}$ and, owing to eq.~(\ref{7}), $B_{\mu\nu}^2\sim 
\frac{g^2}{N}\left<\left(F_{\mu\nu}^a(0)\right)^2\right>$, the neglected terms here 
are of the order of $\left(\frac{g^2\left<\left(F_{\mu\nu}^a(0)\right)^2\right>}{Nm^4}\right)^2$,
i.e., the expansion goes in powers of the parameter  

\begin{equation}
\label{param}
\frac{g^2\left<\left(F_{\mu\nu}^a(0)\right)^2\right>}{Nm^4}.
\end{equation}
For $N=3$, the expansion therefore breaks down for $u$, $d$ and $s$ quarks, but holds for $c$, $b$ and $t$ quarks. Note also that,
as it follows from eq.~(\ref{approx}), 
the range of the $B_{\mu\nu}$-integration in eq.~(\ref{7}) should have been, rigorously speaking, reduced to 
the interval $(-1/T, 1/T)\sim (-m^2, m^2)$. However, the characteristic values of $\sqrt{B_{\mu\nu}^2}$, being of the order of 
${\cal O}\left(\sqrt{\frac{g^2}{N}\left<\left(F_{\mu\nu}^a(0)\right)^2\right>}\right)$, are 
smaller than
$m^2$, as long as the expansion holds. This fact enables us to keep in eq.~(\ref{7}) the infinite range of the $B_{\mu\nu}$-integration with the 
exponential accuracy.

Taking then into account that

$$
\frac{1}{(8\pi C)^3}
\left(\prod\limits_{\mu<\nu}^{}\int\limits_{-\infty}^{+\infty}dB_{\mu\nu}
{\rm e}^{-\frac{B_{\mu\nu}^2}{8C}}\right)\sum\limits_{\alpha<\beta}^{}B_{\alpha\beta}^2=24C,$$
we have:

\begin{equation}
\label{9}
\left<\Gamma\left[A_\mu^a\right]\right>=
-\frac{2NV}{3(4\pi)^2}\cdot 24C
\int\limits_{\Lambda^{-2}}^{\infty}\frac{dT}{T}{\rm e}^{-m^2T}.
\end{equation}
The quark condensate, stemming from this formula by means of eq.~(\ref{1}), reads

\begin{equation}
\label{10}
\left<\bar\psi\psi\right>=-\frac{2NC}{\pi^2m}=-\frac{\alpha_s
\left<\left(F_{\mu\nu}^a(0)\right)^2\right>}{12\pi m},
\end{equation}
that coincides with the result of ref.~\cite{7}. Note that since $\alpha_s\sim\frac1N$ and $\left<
\left(F_{\mu\nu}^a(0)\right)^2\right>\sim N^2$, the condensate~(\ref{10}) grows linearly with $N$,
whereas the parameter~(\ref{param}) is of the order of ${\cal O}\left(N^0\right)$.

Let us now generalize the above-performed analysis
to the case of an arbitrary number $d\ge 2$ of 
space-time dimensions. First, one should take into account that in $d$ dimensions the constant $C$ reads
$\frac{g^2}{8N(d^2-d)}\left<\left(F_{\mu\nu}^a(0)\right)^2\right>$. Further, the field $B_{\mu\nu}$ has 
$n\equiv\frac{d^2-d}{2}$ nontrivial components~\footnote{I.e., a half of all the off-diagonal components.}, 
and the generalization of eqs.~(\ref{7})-(\ref{approx})
takes the form:

\begin{equation}
\label{GG}
\left<\Gamma\left[A_\mu^a\right]\right>\simeq 
-\frac{2NV}{(8\pi C)^{n/2}}\int\limits_{\Lambda^{-2}}^{\infty}\frac{dT}{T}\frac{{\rm e}^{-m^2T}}{(4\pi T)^{d/2}}
\left(\prod\limits_{\mu<\nu}^{}\int\limits_{-\infty}^{+\infty}dB_{\mu\nu}
{\rm e}^{-\frac{B_{\mu\nu}^2}{8C}}\right)\frac{T^2}{3}\sum\limits_{\alpha<\beta}^{}B_{\alpha\beta}^2.
\end{equation}
Clearly, the symbol ``$\simeq$'' here is implied in the sense of the approximation~(\ref{approx}). The latter is controlled 
by the parameter~(\ref{param}), which now acquires the factor $(d^2-d)$ in the denominator, so that the 
accuracy of the approximation enhances in the large-$d$ case. 

Carrying out the $B_{\mu\nu}$-integration, we arrive at the following generalization of 
eq.~(\ref{10})~\footnote{Note that in $d$ dimensions, $[\psi]=m^{\frac{d-1}{2}}$, while $\left[A_\mu^a\right]=
m^{\frac{d}{2}-1}$, $[g]=m^{2-\frac{d}{2}}$, so that $\left[\alpha_s\left<\left(F_{\mu\nu}^a(0)\right)^2\right>\right]=m^4$.}:

\begin{equation}
\label{psge}
\left<\bar\psi\psi\right>=-\frac{m^{d-5}}{3(4\pi)^{\frac{d}{2}-1}}
\alpha_s\left<\left(F_{\mu\nu}^a(0)\right)^2\right>\int\limits_{(m/\Lambda)^2}^{\infty}dtt^{2-\frac{d}{2}}{\rm e}^{-t}.
\end{equation}
This expression diverges as $\ln\frac{\Lambda}{m}$ for $d=6$ and as 
$\Lambda^{d-6}$ for $d>6$, while for $d<6$ it is finite and reads

\begin{equation}
\label{dimens0}
\left<\bar\psi\psi\right>=-\frac{m^{d-5}\Gamma\left(3-\frac{d}{2}\right)}{3(4\pi)^{\frac{d}{2}-1}}
\alpha_s\left<\left(F_{\mu\nu}^a(0)\right)^2\right>
\end{equation}
with ``$\Gamma$'' standing for the Gamma-function. In particular, for $d=2,3$ we obtain

\begin{equation}
\label{dimens}
\left.\left<\bar\psi\psi\right>\right|_{d=2}=-\frac{\alpha_s\left<\left(F_{\mu\nu}^a(0)\right)^2\right>}{3m^3}
\end{equation}
and

\begin{equation}
\label{dIm}
\left.\left<\bar\psi\psi\right>\right|_{d=3}=-\frac{\alpha_s\left<\left(F_{\mu\nu}^a(0)\right)^2\right>}{12m^2}.
\end{equation}

\section{The case 
of the linear form of the next-to-$1/r$ term in the short-distance potential.}
In this section, we shall consider the case when the leading next-to-$1/r$ term 
of the short-distance potential is linear, rather than quadratic. Equation~(\ref{quadr}) is then replaced by

\begin{equation}
\label{14}
\left<W({\cal C})\right>\simeq N\exp\left(-\sigma S_{\rm min}\right)\simeq N\exp\left(-\sigma\sqrt{\frac12
\Sigma_{\mu\nu}^2}\right).
\end{equation}
According to the lattice data~\cite{10}, 

\begin{equation}
\label{siGG}
\sigma=(1\div 5)\sigma_\infty
\end{equation} 
with $\sigma_\infty\simeq (440{\,}{\rm MeV})^2$ is the 
standard string tension at large distances. An analytic expression for $\sigma$ has been obtained in ref.~\cite{yuas}
and reads 

\begin{equation}
\label{siG}
\sigma=\frac{3N\alpha_s\sigma_\infty}{\pi}.
\end{equation}
For the sake of generality, we shall further work directly in the $d$-dimensional
space-time.
Equation~(\ref{14}) can then be rewritten as follows:

$$
\left<W({\cal C})\right>\simeq N\int\limits_{0}^{\infty}\frac{d\lambda}{\sqrt{\pi\lambda}}
\exp\left(-\lambda-\frac{\sigma^2\Sigma_{\mu\nu}^2}{8\lambda}\right)=$$ 

$$=N\int\limits_{0}^{\infty}\frac{d\lambda}{\sqrt{\pi\lambda}}
{\rm e}^{-\lambda}\left(\frac{\lambda}{\pi\sigma^2}\right)^{n/2}
\left[\prod\limits_{\mu<\nu}^{}\int\limits_{-\infty}^{+\infty}dB_{\mu\nu}
\exp\left(-\frac{\lambda}{\sigma^2}B_{\mu\nu}^2\right)\right]
\exp\left(-\frac{i}{2}
B_{\mu\nu}\Sigma_{\mu\nu}\right)=$$

\begin{equation}
\label{longW}
=\frac{\Gamma\left(\frac{n+1}{2}\right)N}{\pi^{\frac{n+1}{2}}\sigma^n}\left(\prod\limits_{\mu<\nu}^{}
\int\limits_{-\infty}^{+\infty}dB_{\mu\nu}\right)
\frac{\exp\left(-\frac{i}{2}
B_{\mu\nu}\Sigma_{\mu\nu}\right)}{\left(1+\frac{1}{2\sigma^2}B_{\mu\nu}^2\right)^{\frac{n+1}{2}}}.
\end{equation}
We are therefore again arriving at the effective action in the constant Abelian field $B_{\mu\nu}$, and only the weight 
of the average over this field is now different from that of eq.~(\ref{GG}). Namely, within the approximation~(\ref{approx}), 
we obtain instead of eq.~(\ref{GG}):

\begin{equation}
\label{ddim}
\left<\Gamma\left[A_\mu^a\right]\right>\simeq 
-\frac{2NV\Gamma\left(\frac{n+1}{2}\right)}{\pi^{\frac{n+1}{2}}\sigma^n}
\int\limits_{\Lambda^{-2}}^{\infty}\frac{dT}{T}\frac{{\rm e}^{-m^2T}}{(4\pi T)^{d/2}}
\left(\prod\limits_{\mu<\nu}^{}\int\limits_{-1/T}^{1/T}dB_{\mu\nu}\right)
\frac{1}{\left(1+\frac{1}{2\sigma^2}B_{\mu\nu}^2\right)^{\frac{n+1}{2}}}
\frac{T^2}{3}\sum\limits_{\alpha<\beta}^{}B_{\alpha\beta}^2.
\end{equation}
Clearly, the symbol ``$\simeq$'' should now be understood in the sense of the smallness of the parameter $\sigma/m^2$, 
rather than the parameter~(\ref{param}). Similarly to the latter, $\sigma/m^2$ is also small for $c$, $b$ and $t$ quarks.

The quark condensate, following from eq.~(\ref{ddim}), reads

\begin{equation}
\label{reads}
\left<\bar\psi\psi\right>=-\frac{N\Gamma\left(\frac{n+1}{2}\right)\sigma^2m^{d-5}}{3\cdot 2^{d-3}\Gamma\left(\frac{n}{2}\right)
\pi^{\frac{d+1}{2}}}\int\limits_{(m/\Lambda)^2}^{\infty}dtt^{2-\frac{d}{2}}{\rm e}^{-t}
\int\limits_{0}^{\frac{m^2}{\sigma t}}\frac{dxx^{n+1}}{\left(1+x^2\right)^{\frac{n+1}{2}}},
\end{equation}
where $x\equiv\frac{1}{\sigma}\sqrt{\frac{B_{\mu\nu}^2}{2}}$.
The dominant contribution to the integral here stems from the region of $t$ around $(m/\Lambda)^2$. For such $t$'s,
the $x$-integral approximately equals $\frac{m^2}{\sigma t}$, and we arrive at the following result: 

\begin{equation}
\label{barpsipsi}
\left.\left<\bar\psi\psi\right>\right|_{d=4}=-\frac{5Nm\sigma}{(4\pi)^2}\ln\frac{\Lambda}{m};
\end{equation}
at $d>4$ the condensate diverges as $(\Lambda/m)^{d-4}$, while at $d<4$ it is finite and equals to

\begin{equation}
\label{cond1}
\left<\bar\psi\psi\right>=-\frac{N\Gamma\left(\frac{n+1}{2}\right)\Gamma\left(2-\frac{d}{2}\right)
\sigma m^{d-3}}{3\cdot 2^{d-3}\Gamma\left(\frac{n}{2}\right)
\pi^{\frac{d+1}{2}}}.
\end{equation}
In particular, 

\begin{equation}
\label{cond2}
\left.\left.\left<\bar\psi\psi\right>\right|_{d=3}=m\left<\bar\psi\psi\right>\right|_{d=2}
=-\frac{2N\sigma}{3\pi^2}.
\end{equation}

Let us now discuss the case $d=4$. The apparent finiteness of the physical condensate in this case implies
that some physical meaning should be attributed to the cutoff $\Lambda$. For usual large-distance 
QCD strings, the latter is the inverse string thickness, i.e., the inverse vacuum correlation length, $T_g^{-1}$~\cite{5}.
It is therefore natural to assume that now $\Lambda$ should be the inverse thickness of a short string, $r_\perp^{-1}$.
Let us also introduce the short-string length $L$, $L=\alpha r_\perp$, where $\alpha\gg 1$.
The two possibilities may then occur for the scale of $L_{\rm min}$: either $L_{\rm min}=\alpha a$, where 
$a=\Lambda^{-1}$ is the lattice spacing (``very short strings''), 
or $L_{\rm min}={\cal O}(T_g)$ (``moderately short strings''). In the former case, the UV cutoff,
$r_\perp^{-1}$, is equal to $a^{-1}$, while in the latter case it equals $\frac{\alpha}{{\cal O}(T_g)}$, that is much larger than $T_g^{-1}$,
but is still much smaller than $a^{-1}$. The case of very short strings is therefore ruled out by the requirement of finiteness of the 
heavy-quark condensate, while moderately short strings may exist. The linear potential is therefore not developed at very short 
distances. In another words, this means that the tachyonic gluon mass vanishes, and the 
perturbative OPE remains valid in the UV region.

Note that
in QCD, to the leading order in the
parameter of the strong-coupling expansion, $\beta_{\rm sc}=2N/g^2$, the string tension 
is known to be $N$-independent. Nemely, $\sigma_{\infty}=\frac{1}{a^2}\ln\frac{2N^2}{\beta_{\rm sc}}=\frac{1}{a^2}\ln\lambda$,
where $\lambda=g^2N$ is the 't~Hooft coupling~\footnote{The $N$-independence 
of $\sigma_{\infty}$ stems also from the natural conjecture that the linear term in the
large-distance quark-antiquark potential should have the same $N$-dependence as the Coulomb term, that is
$V_{\rm Coul}(r)=-\frac{g^2}{4\pi r}\frac{N^2-1}{2N}={\cal O}\left(N^0\right)$.}. Therefore, as it follows from eqs.~(\ref{siGG}) and 
(\ref{siG}),
to the leading order in $\beta_{\rm sc}$, 
$\sigma$ is $N$-independent as well,
and the expressions~(\ref{barpsipsi}), (\ref{cond2}) 
grow linearly with $N$ in the same way as the expressions~(\ref{10}), (\ref{dimens}), (\ref{dIm}).

\section{Finite-temperature generalizations in four dimensions.}

In this section, we shall generalize the above-obtained 4d-results~(\ref{10}) and~(\ref{barpsipsi})~\footnote{ 
In the latter case, as it has been discussed above, it is implied that $\Lambda$ is attributed the physical meaning of the 
inverse short-string thickness.} to the case of finite temperature~\footnote{It should not be confused
with the proper time! To avoid possible confusions, we shall frequently use
below the inverse temperature, $\beta$.}.
In the case of eq.~(\ref{10}), the main ingredient necessary for such a generalization 
is the formula which properly accounts for the antiperiodic boundary conditions for quarks, one should
take care of in the course of integration over them at finite temperature. Such a formula has been derived 
in ref.~\cite{8}, where it has been shown that at finite temperature, the (zero-temperature) heat
kernel for fermions becomes multiplied by
the factor 

\begin{equation}
\label{bc}
1+2\sum\limits_{n=1}^{\infty}(-1)^n{\rm e}^{-\frac{\beta^2n^2}{4T}},
\end{equation}
where $n$ labels the Matsubara mode. 
We therefore obtain from eq.~(\ref{9}):

$$
\left<\bar\psi\psi\right>=
-\frac{2N}{3(4\pi)^2}\cdot 24C\left(-\frac{\partial}{\partial m}\right)
\int\limits_{\Lambda^{-2}}^{\infty}\frac{dT}{T}{\rm e}^{-m^2T}\left[
1+2\sum\limits_{n=1}^{\infty}(-1)^n{\rm e}^{-\frac{\beta^2n^2}{4T}}\right]=$$

\begin{equation}
\label{13}
=\left<\bar\psi\psi\right>_0
\left[1+2\beta m\sum\limits_{n=1}^{\infty}(-1)^nnK_1(\beta mn)\right],
\end{equation}
where $K_\nu$'s from now on stand for the modified Bessel functions, and
$\left<\bar\psi\psi\right>_0$
is given by eq.~(\ref{10}) with the temperature-dependent gluonic condensate. The latter is known to 
experience a drop by a factor of the order 2 at the deconfinement temperature, $T_c$, due to the evaporation of the 
chromoelectric condensate~\cite{Temp}. For $c$, $b$ and $t$ quarks under discussion,
the QCD deconfinement temperature is much smaller than the quark mass, i.e., $\beta m\gg 1$. 
Within this inequality,
only the first Matsubara mode yields a sufficient 
contribution to the sum in eq.~(\ref{13}), whereas all higher modes may be disregarded. In this case, 
we obtain the following leading (exponentially small) correction to the heavy-quark condensate:

\begin{equation}
\label{lc}
\left<\bar\psi\psi\right>\simeq\left<\bar\psi\psi\right>_0
\left(1-\sqrt{2\pi\beta m}{\rm e}^{-\beta m}\right).
\end{equation}

When the temperature $T$ is larger than
the temperature of dimensional reduction in QCD, $T_{\rm d.r.}$,
the theory becomes three-dimensional. Consequently, 
eqs.~(\ref{13}), (\ref{lc}) are no more valid, and the derivation should rather be based on eq.~(\ref{psge}) at $d=3$.
Using eq.~(\ref{bc}) and performing the proper-time integration, we obtain instead of eq.~(\ref{13}):

$$
\left<\bar\psi\psi\right>
=\left<\bar\psi\psi\right>_0
\left[1+\frac{(2\beta m)^{3/2}}{\sqrt{\pi}}
\sum\limits_{n=1}^{\infty}(-1)^nn^{3/2}K_{3/2}(\beta mn)\right],
$$
where $\left<\bar\psi\psi\right>_0$ is given by eq.~(\ref{dIm})~\footnote{Clearly, at such temperatures, the gluonic condensate 
in eq.~(\ref{dIm}) is purely chromomagnetic.}.
Since $T_{\rm d.r.}$ does not exceed $2T_c$~\cite{ba2}, 
for $c$, $b$, and $t$ quarks under discussion a rather broad range of temperatures above $T_{\rm d.r.}$ exists, in which $\beta m\gg 1$.
For such temperatures, we obtain instead of eq.~(\ref{lc}):

$$
\left<\bar\psi\psi\right>\simeq\left<\bar\psi\psi\right>_0
\left(1-2\beta m{\rm e}^{-\beta m}\right),
$$
where $\left<\bar\psi\psi\right>_0$ is again given by eq.~(\ref{dIm}).

Let us now consider the finite-temperature generalization of eq.~(\ref{barpsipsi}). Note that in this case, since 
$r_\perp$ grows with the temperature,
one may substitute $r_\perp$ for $\Lambda^{-1}$ only for such temperatures, at which the inequality $mr_\perp(T)\ll 1$ holds.
For the sake of generality, let us assume that this is true up to temperatures larger than $T_{\rm d.r.}$. Then,
at $T<T_c$, the direct combination of eq.~(\ref{reads}) (at $d=4$) with eq.~(\ref{bc}) yields

\begin{equation}
\label{tmt}
\left<\bar\psi\psi\right>
=\left<\bar\psi\psi\right>_0
\left[1+\frac{2}{\ln\frac{\Lambda}{m}}
\sum\limits_{n=1}^{\infty}(-1)^nK_0(\beta mn)\right]\simeq
\left<\bar\psi\psi\right>_0\left(1-\sqrt{\frac{2\pi}{\beta m}}\frac{{\rm e}^{-\beta m}}{\ln\frac{\Lambda}{m}}\right),
\end{equation}
where $\left<\bar\psi\psi\right>_0$ is given by eq.~(\ref{barpsipsi}). 
At $T>T_c$,
only the spatial part of the string tension, $\sigma_s$, does not vanish,
and the Wilson loop reads [cf. eqs.~(\ref{14}), (\ref{longW})]:

\begin{equation}
\label{BMU}
\left<W({\cal C})\right>\simeq N\exp\left[-\sigma_s\sqrt{\Sigma_{12}^2+\Sigma_{13}^2+\Sigma_{23}^2}\right]=
\frac{N}{\pi^2\sigma_s^3}\left(\prod\limits_{\mu<\nu}^{}
\int\limits_{-\infty}^{+\infty}dB_{\mu\nu}\right)
\frac{\exp\left(-\frac{i}{2}
B_{\mu\nu}\Sigma_{\mu\nu}\right)}{\left(1+\frac{1}{2\sigma_s^2}B_{\mu\nu}^2\right)^2}.
\end{equation}
One should further distinguish the cases $T\in\left[T_c, T_{\rm d.r.}\right]$ and $T>T_{\rm d.r.}$. In the former,  
although $B_{\mu\nu}$ in eq.~(\ref{BMU}) is a $3\times 3$-matrix, the theory is still four-dimensional, and one should therefore 
substitute $d=4$ into the factor $(4\pi T)^{-d/2}$ of the Euler-Heisenberg-Schwinger formula on the r.h.s. of eq.~(\ref{ddim}).
This yields

$$
\left<\bar\psi\psi\right>
=-\frac{2Nm\sigma_s}{3\pi^3}\ln\frac{\Lambda}{m}
\left[1+\frac{2}{\ln\frac{\Lambda}{m}}
\sum\limits_{n=1}^{\infty}(-1)^nK_0(\beta mn)\right]\simeq
-\frac{2Nm\sigma_s}{3\pi^3}\ln\frac{\Lambda}{m}
\left(1-\sqrt{\frac{2\pi}{\beta m}}\frac{{\rm e}^{-\beta m}}{\ln\frac{\Lambda}{m}}\right).
$$
The difference of the overall factor $\frac{2Nm\sigma_s}{3\pi^3}$ in this expression from the factor
$\frac{5Nm\sigma}{(4\pi)^2}$ of eq.~(\ref{tmt}) is clearly due to the fact that the temporal part of the Wilson loop
vanishes when one passes from $T<T_c$ to $T>T_c$. Note that, naively, one could have expected at $T=T_c+0$ the factor 
$\frac{5Nm\sigma_s}{(4\pi)^2}$, that turned out not to be the case. However, the numerical difference of the two
factors is small, since $\frac{2}{3\pi}\simeq0.21$, while $\frac{5}{16}\simeq0.31$.
Finally, at $T>T_{\rm d.r.}$, the theory becomes fully three-dimensional, and we obtain from eq.~(\ref{reads}) at $d=3$
[combined with eq.~(\ref{bc})]

$$
\left<\bar\psi\psi\right>
=\left<\bar\psi\psi\right>_0
\left[1+2\sqrt{\frac{2\beta m}{\pi}}
\sum\limits_{n=1}^{\infty}(-1)^n\sqrt{n}K_{1/2}(\beta mn)\right]\simeq
\left<\bar\psi\psi\right>_0\left(1-2{\rm e}^{-\beta m}\right).$$
Here, $\left<\bar\psi\psi\right>_0$ is given by eq.~(\ref{cond2}) with the substitution $\sigma\to\sigma_s$.

\section{Summary.}

In the present letter, we have applied the world-line formalism to a derivation of the heavy-quark condensate in QCD.
The main idea implemented is that, due to the fact that the heavy-quark trajectories are quite small, the respective Wilson 
loop can be expressed in terms of the so-called tensor area. The resulting path-integral then reduces to the one in an auxiliary
constant Abelian field, that is a famous Euler-Heisenberg-Schwinger Lagrangian. This auxiliary field
should then be averaged over with a weight prescribed by the form of the short-distance quark-antiquark
potential. Such an approximation of the Wilson loop by a function of the tensor area has been first successfully tested in section 2.
There, the quadratic form of the next-to-$1/r$ term of the potential, which follows from the VCM, has been adopted, and as a result
the QCD-sum-rules expression for the heavy-quark condensate in four dimensions has been reproduced. This result has further readily been 
generalized to the case of an arbitrary number $d\ge 2$ of space-time dimensions. (In particular, it has been found that at $d\ge 6$, the 
condensate is UV divergent.) Being encouraged with this successful test, we have proceeded in section 3 with another tensor-area-based
Ansatz for a small Wilson loop, namely the one which produces the linear short-distance potential. Surprisingly, we have found that 
in this case the condensate is explicitly UV finite only at $d<4$, whereas at $d=4$ it is 
(logarithmically) divergent. We have further argued that the problem of this divergency can be bypassed (by attributing 
to the UV cutoff the physical meaning of the inverse string thickness) only provided 
very short strings (of a length comparable with the lattice spacing) are forbidden. This means that the tachyonic gluon mass,
which produces the short-distance linear potential, should vanish in the UV region, and the perturbative OPE should remain valid 
in this region. Such a conclusion coincides with the one of ref.~\cite{hrs}, where it has been drawn from the direct evaluation of the 
tachyonic gluon mass. 

We have also demonstrated that the $N$-dependence of the condensate corresponding to the linear form of the 
next-to-$1/r$ term in the short-distance potential is the same as the one of the standard condensate corresponding to the quadratic
term, namely both condensates grow linearly with $N$.
We have further studied the finite-temperature behaviors of both condensates.
Specifically, it turned out that for heavy (i.e., $c$, $b$, and $t$) quarks under study, accounting for antiperiodic boundary
conditions for fermions leads to corrections, which are exponentially small in the parameter $\beta m$ (both below the temperature of dimensional
reduction and in the broad range of temperatures above it). As for the leading term, in the case of quadratic short-distance 
potential it is given by the corresponding $4d$- and $3d$-expressions below and above the temperature of dimensional reduction, respectively,
with the gluonic condensate saturated above the deconfinement temperature by its chromomagnetic part only. In the case of the linear
potential, the whole analysis is legitimate only within such temperatures, at which the short-string thickness, which grows with the temperature,
remains much smaller than the inverse quark mass. Assuming the general case when this is true up to temperatures higher than the temperature
of dimensional reduction, we have demonstrated that the leading term of the quark condensate is different in the region between the 
deconfinement temperature and the temperature of dimensional reduction, and in the region of temperatures higher than the temperature
of dimensional reduction.
Namely, while in the latter region this term is merely the zero-temperature $3d$-result with the full string tension replaced by the 
spatial one, in the former region the result is completely novel. That is because in this region the theory is still four-dimensional,
but the quark Wilson loop is that of the $3d$-theory (since only the spatial string tension survives the deconfinement phase transition).
As a consequence, only the parametric dependence of the respective result on $N$, $m$, and $\sigma_s$ 
is the same as that of the zero-temperature $4d$-one (with the 
spatial string tension replaced by the full one), while the numerical factors of the two results are different, albeit close
to each other.

\section{Acknowledgments.}

The author is grateful to Profs. A.~Di~Giacomo and 
Yu.A.~Simonov for very useful discussions.
He has also benefited from discussions with Profs.
H.G.~Dosch, D.~Ebert, F.~Lenz and Yu.M.~Makeenko, and Drs. C.~Schubert and K.~Zarembo.
Besides that, the author is grateful to
the whole staff of the Physics Department of the
University of Pisa for cordial hospitality.
The work has been supported by INFN.

\end{document}